%% file: main.tex
\documentclass[conference]{IEEEtran}
\IEEEoverridecommandlockouts
\usepackage{cite}
\usepackage{amsmath,amssymb,amsfonts}
\usepackage{algorithmic}
\usepackage{graphicx}
\usepackage{textcomp}
\usepackage{xcolor}
\usepackage{url}
\def\BibTeX{{\rm B\kern-.05em{\sc i\kern-.025em b}\kern-.08em
    T\kern-.1667em\lower.7ex\hbox{E}\kern-.125emX}}
\begin{document}

\title{Data Distribution Dynamics in Real-World WiFi-Based Patient Activity Monitoring for Home Healthcare}

\author{Mahathir Monjur$^1$, Jia Liu$^3$, Jingye Xu$^2$, Yuntong Zhang$^2$, Xiaomeng Wang$^3$, Chengdong Li$^4$, Hyejin Park$^4$, \\ Wei Wang$^2$,
Karl Shieh$^5$, Sirajum Munir$^6$, Jing Wang$^4$, Lixin Song$^3$, Shahriar Nirjon$^1$\\

\{mahathir, nirjon\}@cs.unc.edu, \{cli7, hpark5\}@fsu.edu, \{jingye.xu, yuntong.zhang, wei.wang\}@utsa.edu \\

\{liuj14, wangx14, songl2\}@uthscsa.edu, kshieh12@ad.unc.edu, jingwang@nursing.fsu.edu, mus1pi@bosch.com\\

$^1$Department of Computer Science, University of North Carolina at Chapel Hill, NC\\
$^2$Department of Computer Science, University of Texas at San Antonio, TX\\
$^3$School of Nursing, University of Texas Health Science Center at San Antonio, TX\\  
$^4$College of Nursing, Florida State University, FL\\
$^5$School of Nursing, University of North Carolina at Chapel Hill, NC\\
$^6$Bosch Research and Technology Center, Pittsburgh, PA\\
}

\newcommand{\Sys}{AURA}
\newcommand\rnote[1]{\textcolor{red}{#1}}
\newcommand\gnote[1]{\textcolor{green}{#1}}
\newcommand\bnote[1]{\textcolor{blue}{#1}}
\newcommand\pnote[1]{\textcolor{purple}{#1}}

\maketitle

\begin{abstract}

This paper examines the application of WiFi signals for real-world monitoring of daily activities in home healthcare scenarios. While the state-of-the-art of WiFi-based activity recognition is promising in lab environments, challenges arise in real-world settings due to environmental, subject, and system configuration variables, affecting accuracy and adaptability. The research involved deploying systems in various settings and analyzing data shifts. It aims to guide realistic development of robust, context-aware WiFi sensing systems for elderly care. The findings suggest a shift in WiFi-based activity sensing, bridging the gap between academic research and practical applications, enhancing life quality through technology.

\end{abstract}


\input{tex/01_introduction}
\input{tex/02_system}

\input{tex/03_dataset}
\input{tex/04_analysis}
\input{tex/05_discussion}
\input{tex/06_conclusion}

%
%\begin{acks}
%This work was supported, in part, by grants NIH 5R01LM013329-02 and NSF 2047461.
%\end{acks}
%

%\bibliographystyle{ACM-Reference-Format}
% \bibliography{sample}
% \begin{thebibliography}{00}
% \bibitem{b1} Chen, Z., Zhang, L., Jiang, C., Cao, Z. and Cui, W., 2018. WiFi CSI based passive human activity recognition using attention based BLSTM. IEEE Transactions on Mobile Computing, 18(11), pp.2714-2724.
% \bibitem{b2}Wang, W., Liu, A.X., Shahzad, M., Ling, K. and Lu, S., 2015, September. Understanding and modeling of wifi signal based human activity recognition. In Proceedings of the 21st annual international conference on mobile computing and networking (pp. 65-76).
% \bibitem{b3}Schulz, M., Wegemer, D. and Hollick, M., 2017, October. Nexmon: Build your own wi-fi testbeds with low-level mac and phy-access using firmware patches on off-the-shelf mobile devices. In Proceedings of the 11th Workshop on Wireless Network Testbeds, Experimental evaluation \& CHaracterization (pp. 59-66).
% \end{thebibliography}

\bibliographystyle{abbrv}
\bibliography{main}

\end{document}

%% file: tex/01_introduction.tex
\section{Introduction}

WiFi signals, a common household technology, can be repurposed for detecting and classifying daily activities in innovative ways. These signals, emitted by routers, permeate rooms and interact with objects and people within. When a person moves or performs activities, they disrupt the WiFi signal pattern. Sensing systems detect these disruptions, interpreting them as specific activities like walking or sitting. By analyzing signal strength changes and patterns, the system can accurately identify different movements. This technology, therefore, transforms ordinary WiFi into a tool for monitoring daily life unobtrusively.

In healthcare domain, WiFi sensing is gaining traction due to its ability to monitor various activities and health-related events without physical contact or the need for wearable devices~\cite{chen2018wifi, zou2017multiple, wang2015understanding, ding2019wifi, shang2021lstm}. This technology can detect critical events like falls, sleep disturbances, wandering behavior, respiratory disorders, and abnormal cardiac activity, particularly in vulnerable populations such as the elderly. Its appeal lies in its ease of operation indoors and the comfort it provides to individuals being monitored, as it is less intrusive compared to wearables \cite{zhuang2019design}, RGBD sensors \cite{zhang2012rgb}, camera images \cite{tan2018multi}, or acoustic solutions \cite{sim2015acoustic}. 

Unfortunately, while WiFi sensing in healthcare shows promise in academic studies, it encounters substantial difficulties transitioning from the controlled settings of a lab to the complexities of real-world environments. In laboratories, conditions are ideal, with minimal interference and predictable variables. %This allows for the fine-tuning of systems to detect specific activities or health events with high accuracy.% 
However, in uncontrolled environments like homes, where patients reside for home healthcare or aging in place, numerous unpredictable factors come into play. Factors such as varying WiFi signal strengths, different home layouts, presence of multiple individuals, and diverse daily activities create a complex environment. This complexity often leads to decreased accuracy and reliability of WiFi sensing systems. They struggle to adapt to the dynamic and individualized nature of home settings, where conditions differ significantly from the lab. This gap between lab-controlled conditions and real-world variability is a critical limitation that needs addressing for effective deployment in healthcare scenarios.

In this paper, we share our experience from a multi-year study where we deployed WiFi sensing systems in 8 different environments with 16 participants involved. These environments encompass a variety of settings such as offices, kitchens, laboratories, and living room areas, while the participant demographic is diverse, including males and females, young and elderly, and both healthy and  weak patients. Additionally, before these deployments, we had trained and tested the system in various lab environments. Our deployments varied in duration, ranging from a few hours to a maximum of two weeks of continuous monitoring. In each deployment, we recorded the ground truth using cameras and/or wearable devices. From the data we collected, we observed significant shifts in data distribution due to multiple factors. We summarize our preliminary findings through a comprehensive study involving:

\begin{itemize}
    
    \item We analyze various factors such as individual presence, environmental conditions, and time intervals affecting WiFi signal variability. This detailed analysis aids in understanding and quantifying the contribution of each factor to the overall shift in data distribution.
    
    \item Our rigorously evaluate existing data adaptation techniques, assessing their effectiveness beyond the controlled settings of a lab.
    
    \item Utilizing machine learning and signal processing algorithms, we aim to uncover hidden patterns within the chaotic data, providing insights into the mechanisms driving dataset shift.
\end{itemize}

This research will significantly advance elderly care monitoring by developing adaptive sensing systems, leading to a paradigm shift in WiFi-based activity sensing. It aims to unify academic and industrial efforts, enhancing quality of life through innovative and context-aware technology.

%% file: tex/02_system.tex
\section{AURA System}

\begin{figure}[!htb]
    \includegraphics[width=0.49\textwidth]{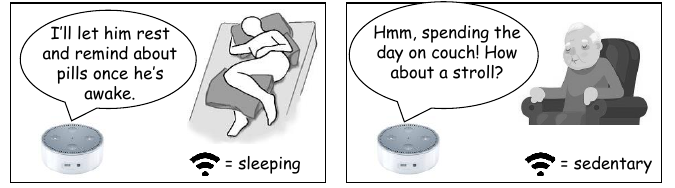}
    \caption{User's activity-aware conversational assistant.}
    \label{fig:aura_cartoons}
\end{figure}

This paper is part of a multi-year NIH project, \emph{``AURA,''}, which enhances voice assistants like Alexa with WiFi sensing for home healthcare. Figure~\ref{fig:aura_cartoons} shows AURA's use in detecting activities, enabling smarter, context-aware Alexa interactions.  

An AURA-enabled voice assistant can monitor users and their environment passively. This means it operates without needing frequent user interactions, which can be bothersome, or requiring the user to carry extra devices such as GPS or health trackers, which are often forgotten or inconvenient. Importantly, the AURA system avoids using cameras to respect privacy, a crucial factor in home environments where both patients and their families are present. WiFi sensing, unlike cameras, functions in darkness and can penetrate walls, enabling AURA to extend its reach beyond the confines of a single room. However, for calibration and ground truth recording purposes, cameras were temporarily employed during the initial hour of our study.

\begin{figure}
    \includegraphics[width=0.47\textwidth]{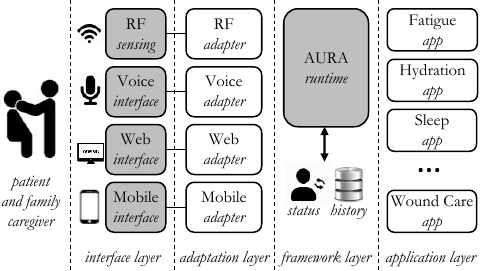}
    \caption{Architecture of AURA.}
    \label{fig:aura_arch}
    \vspace{-10pt}
\end{figure}

%\begin{figure}[!htb]
%\includegraphics[width=0.9\linewidth]{figs/system_crop_updated.pdf}
%    \caption{Architecture of AURA.}
%    \label{fig:aura_arch}
%    \vspace{-10pt}
%\end{figure}

The architecture of AURA is depicted in Figure~\ref{fig:aura_arch}. AURA offers four interaction interfaces for patients and caregivers: WiFi (RF), voice, web, mobile. Data from these interfaces are standardized by adapters before entering the AURA runtime. This design ensures easy extension with new interactions without altering the core system. The AURA runtime monitors patient status, logs data, offers APIs, and manages AURA apps execution. The number and types of apps vary with the deployment scenario. Each AURA app monitors patient needs specific to their health application, such as hydration and sleep for post-treatment CBC patients. Using AURA's APIs, these apps convert WiFi sensing data, like movement frequency and drinking gestures, into clinically relevant information.

In this paper, we focus solely on AURA's WiFi data collection, analysis, and activity classifier modules.

\begin{figure*}[!htb]
    \includegraphics[width=\textwidth]{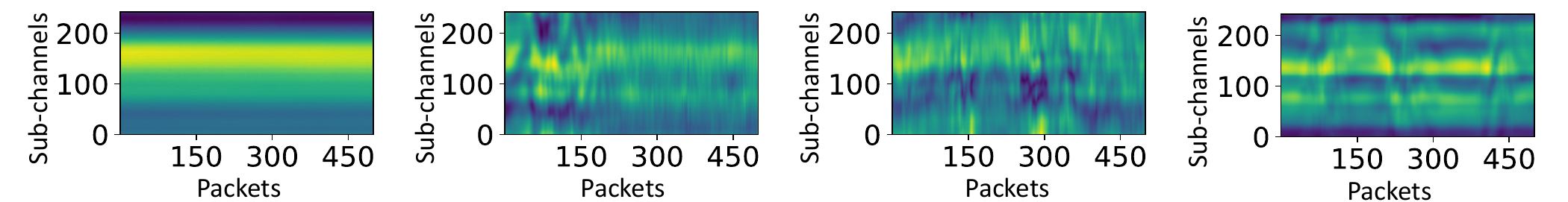}
    \caption{Processed WiFi CSI for (a)~no activity, (b)~fall, (c)~jogging, and (d)~squat.}
    \label{fig:example}
    \vspace{-10pt}
\end{figure*}

%% file: tex/03_dataset.tex
\section{Dataset}
\subsection{Study Design}

In this study, we conducted several recording sessions involving 16 participants. Before each session commenced, we briefed participants on the activities they were to perform and informed them about our data handling procedures, including anonymization and publication policies. We also obtained written consent from all participants.

Each session involved collecting a paired dataset consisting of image sequences (RGB and depth) and WiFi CSI. To capture the RGB and depth images, we used an Intel D435 depth camera, set to \(1280 \times 720\) resolution and 10 fps. For CSI data, we employed a pair of ASUS RT-AC86U routers. One router was configured to transmit packets at 200 packets per second rate on a single spatial channel. The other router was linked to an NVIDIA Jetson Nano computer to receive the packets. Both routers were set to monitor mode using the Nexmon framework~\cite{schulz2017nexmon}. To ensure synchronization, the collection of WiFi CSI data and camera image data was initiated concurrently via the jetson nano computer and each camera image frame was timestamped. The WiFi packets also contained microsecond-level time information, which was later used to align each camera and WiFi CSI data pair.

We collected paired image and CSI data from 16 participants across 8 distinct environments. During each session, we continuously recorded camera and WiFi CSI data while participants engaged in various activities. These activities were annotated based on their start and end times, corresponding to specific image frames, which enabled us to extract the relevant WiFi CSI packets. These sessions resulted in a comprehensive dataset of 606 minutes or \~10 hours. The participant group comprised 7 males and 9 females. In addition to the paired data, we also gathered WiFi CSI data specific to different human activities (with labels) from 3 participants in 5 different environments.

\subsection{Data Processing}
After collecting raw WiFi CSI data, we arrange the packets by the number of antennas and then apply fft on the raw data. This results in a spectrogram with the shape of $(N\times N_{ant})\times M$, where $N$ is number of packets, $N_{ant}$ is number of antenna used from the receiver router which can be $1$ or $4$, and $M$ is the number of sub-channels. Although our setup was configured for collecting data on $80$ MHz bandwidth and $256$ sub-channels, Nexmon allows CSI collection of $M=242$ data-channels while the other $14$ channels is used as control sub-channels. Hence we remove the additional channels with no information. Examples of processed CSI data for different activities are shown in Figure~\ref{fig:example}.

\subsection{Patient Activity Monitoring}
In our dataset, participants engaged in a diverse array of activities such as walking, eating, drinking, light arm exercises, jogging, squat, jumping, falling, sitting down, standing up, and remaining seated or standing with minimal movement. These activities were continuously recorded using both cameras and WiFi routers using single antenna. The sessions were meticulously labeled post-recording, with the start and end image frames for each activity sequence being marked to create a comprehensive metadata file. Additionally, we have augmented our original dataset from patients with labeled CSI data, encompassing 4 channels, for three extra participants across five different environments. A detailed statistics of the paired image and WiFi dataset is shown in~\ref{tab1}, where we list each unique environment, and the number of male and female participants, number of image frames, and number of CSI packets collected within that environment. In total, we collected 232,440 image frames and 7,436,828 CSI packets. This enriched dataset provides a robust foundation for detailed activity recognition research and analysis. The dataset and relevant scripts can be found here: \url{https://github.com/Monjur-Mahathir/AURA-DNN.git}.

\begin{table}[htbp]
\caption{Dataset Statistics}
\begin{center}
\vspace{-5pt}
\begin{tabular}{|c|c|c|c|}
\hline 
\textbf{Environment}& \textbf{\# of participants} & \textbf{\# of images} & \textbf{\# of CSI packets} \\
\hline
Living Room-1 & 1 Male & 21,779 & 399,687\\
\hline
Living Room-2 & 1 Female & 11,991 & 429,607\\
\hline
Living Room-3 & 1 Male & 20,210 & 458,948\\
\hline
Living Room-4 & 1 Female & 11,000 & 413,181\\
\hline
Living Room-5 & 1 Male & 7,707 & 389,985\\
\hline
Living Room-6 & 1 Male, 1 Female & 20,986 & 1,065,504\\
%Kitchen-4 & 4 Female, 3 Male & 10,242+11278+10989+13004+24933+11216+11216+ & %486,968+533558+529167+656969+570570+560775+500000+\\
\hline
Kitchen-7 & 4 Female, 3 Male & 92,878 & 3,838,007\\
\hline
Office-8 & 2 Female & 45,889 & 441,909\\
\hline
\end{tabular}
\label{tab1}
\end{center}
\end{table}

\begin{figure*}[!htb]
    \begin{center}
        \includegraphics[width=0.82\textwidth]{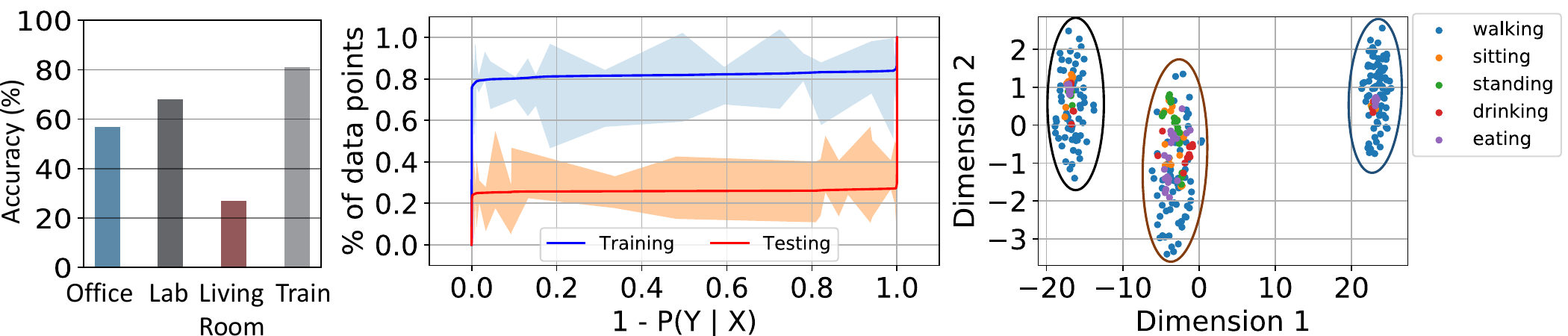}
    \end{center}
    \vspace{-10pt}
    \caption{Data shift due to environment change.}
    \label{fig:env_effect}
    \vspace{-10pt}
\end{figure*}

\begin{figure*}[!htb]
   \begin{center}
        \includegraphics[width=0.82\textwidth]{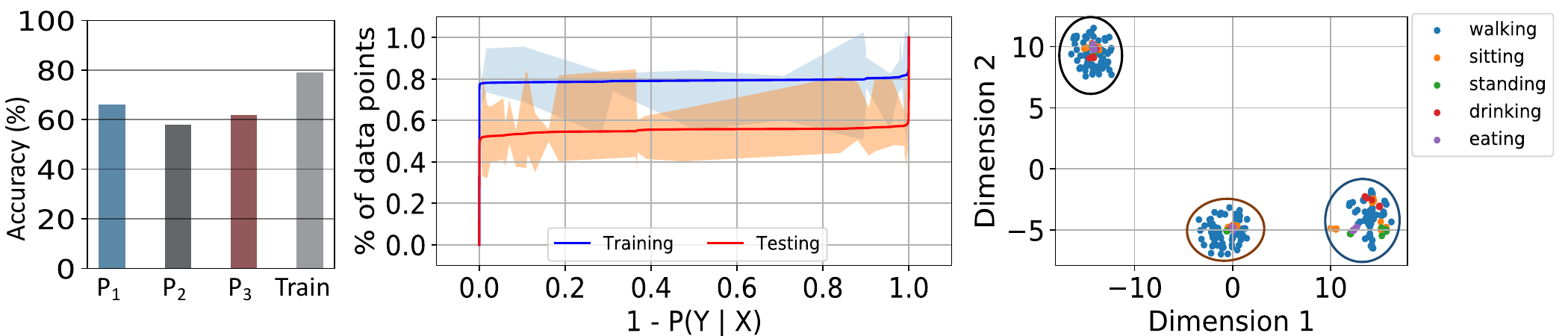}
    \end{center}
    \vspace{-10pt}
    \caption{Data shift due to activity from different people.}
    \label{fig:people_effect}
    \vspace{-10pt}
\end{figure*}

%% file: tex/04_analysis.tex
\section{Dataset Analysis}

\subsection{Data Analysis with t-SNE Embedding}
\begin{figure}[!b]
    \vspace{-10pt}
    \includegraphics[width=0.47\textwidth]{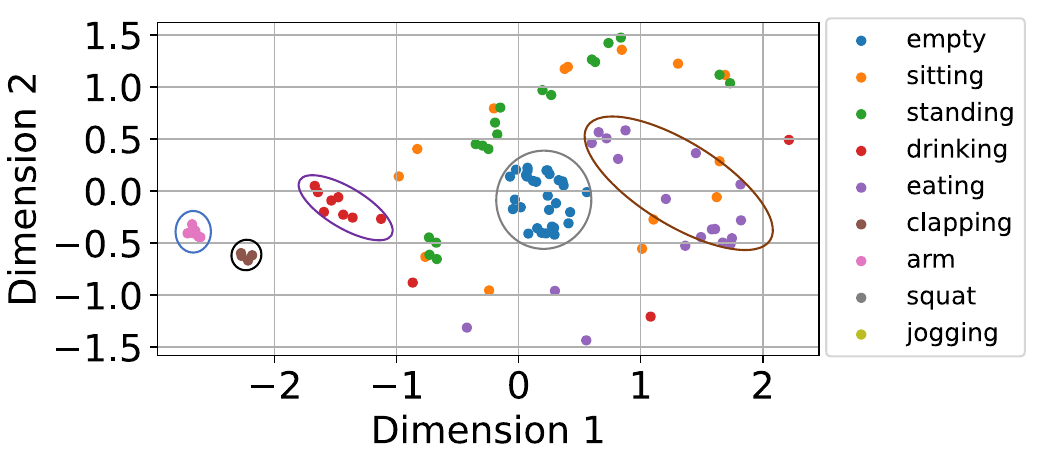}
    \caption{t-SNE embedding of WiFi CSI data}
    \label{fig:tsne}
\end{figure}

To analyze the effectiveness of the WiFi CSI data collected from different patient's home, we used t-Distributed stochastic neighbor embedding~(t-SNE), which is a powerful technique for visualizing high-dimensional data by reducing it to lower-dimensional spaces, while maintaining the local structure of the data. It starts by converting the distances between data points in high-dimensional space into conditional probabilities that represent similarities. The similarity of datapoint \( x_j \) to datapoint \( x_i \) is given by a conditional probability \( p_{j|i} \), which is computed as:

\[ p_{j|i} = \frac{\exp(-||x_i - x_j||^2 / 2\sigma_i^2)}{\sum_{k \neq i}\exp(-||x_i - x_k||^2 / 2\sigma_i^2)} \]

where \( ||x_i - x_j|| \) is the Euclidean distance between \( x_i \) and \( x_j \), and \( \sigma_i \) is the variance of the Gaussian that is centered on datapoint \( x_i \). In the low-dimensional space, a similar probability \( q_{j|i} \) is calculated, often using a Student’s t-distribution. The aim of t-SNE is to minimize the divergence between these two probability distributions, typically using the Kullback-Leibler divergence as a cost function. This minimization is usually achieved through gradient descent. The result is a map that reveals the intrinsic structure of the data in a way that is easily interpretable, often revealing clusters and relationships that are not apparent in the high-dimensional space. In Figure~\ref{fig:tsne}, we can see that the raw CSI data from a session forms several differentiable clusters for different activities.

\subsection{Dataset Shift}

\begin{figure*}[!htb]
    \begin{center}
        \includegraphics[width=0.82\textwidth]{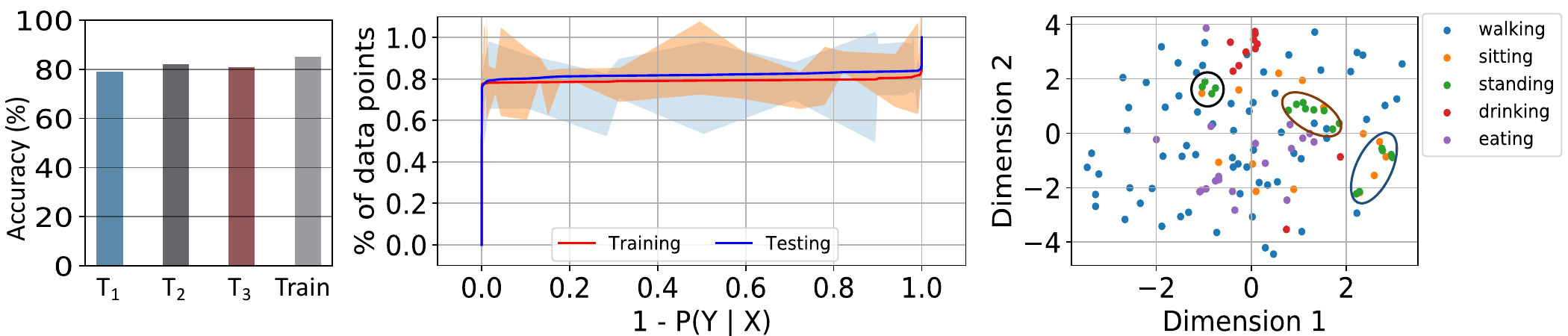}
    \end{center}
    \vspace{-10pt}
    \caption{Data shift due to temporal variation.}
    \label{fig:time}
    \vspace{-10pt}
\end{figure*}

\subsubsection{Dataset Shift from Environment}\label{sec:shift_e}
To evaluate the adaptability of WiFi CSI data in patient activity monitoring across new environments, we partition our dataset into training and testing subsets. Notably, the testing set comprises data from three entirely new environments: an office, a lab, and a living room. After training our DNN with the training set, we test its performance on this unseen data. As depicted in Figure~\ref{fig:env_effect}, the model's accuracy notably decreases (ranging between 13\%-54\%) when applied to these new environments. This drop in performance is particularly pronounced in the complex and object-filled environment of the living room, while it is less severe in more controlled settings like the lab, which has fewer objects.

We further analyze the shift in performance by comparing the softmax probability distributions of the outputs on the training set, \( P_{train}(y|X) \), and the testing set from the unseen environments, \( P_{test}(y|X) \). This comparison reveals significant differences in the model's behavior between familiar and unfamiliar settings. In addition, a t-SNE visualization, which projects the high-dimensional feature space generated by the DNN, \( Z = f_{\theta}(X) \), into a two-dimensional space for enhanced interpretability, demonstrates that the data from these three distinct environments form three separate, clearly identifiable clusters. This clustering, represented in the t-SNE plot, is indicative of the varying environmental impacts on the model's performance.

\subsubsection{Dataset Shift from People}
In our study to assess dataset shift in CSI data for activity recognition involving unseen participants, we conduct an experiment where three individuals from a single environment are intentionally excluded from the training dataset, while data from two different individuals from the same environment are retained in the training set. The results show a moderate accuracy degradation of 12-20\%, a contrast to the more substantial drops observed in Section~\ref{sec:shift_e} for completely unseen environments. Further analysis, including the probability distribution and t-SNE plots presented in Figure~\ref{fig:people_effect}, reveals a relatively minor data shift when different participants were involved in a familiar environmental setting, underscoring the impact of environmental familiarity on the model's performance in activity recognition tasks.

\subsubsection{Dataset Shift from Time}
In our investigation to determine the temporal stability of WiFi CSI data and its influence on dataset distribution, we conduct an analysis involving three distinct sessions. We strategically partition the data from these sessions, incorporating the first half into the training set and the latter half into the testing set. This approach aims to observe potential changes in the probability distribution over time. The results, as depicted in Figure~\ref{fig:time}, indicates a minimal impact of dataset shift over time, provided that both the environment and the participants remain consistent within the training set. This finding suggests that temporal variations do not significantly affect the dataset distribution for activity recognition tasks when the environmental and participant variables remain constant, thereby affirming the temporal robustness of the CSI data under controlled conditions.

%% file: tex/05_discussion.tex
\section{Activity Recognition}

\subsection{Model Description}
To facilitate the deployment of our DNN on resource-constrained NVIDIA Jetson Nano, we engineer a compact and computationally efficient DNN architecture. This architecture integrates a sequence of three convolutional layers, each succeeded by a batch normalization layer and a max pooling operation. Following the convolutional stack, the processed feature set is input into three stacked bidirectional LSTM layers, designed to capture temporal dependencies effectively. The output at the final timestep of these LSTM layers is then channeled into a dense layer, which is coupled with a softmax activation function to facilitate multi-class classification. The DNN's total size is approximately 6.761MB, encompassing around 1.77 million parameters.

In our baseline experiment, the DNN was trained over 500 epochs using an Adam optimizer with a learning rate of 0.001 and a batch size of 16, with the dataset divided into training and testing sets in an 80-20\% split. To address domain-specific effects, we extended our analysis to include two domain adaptation strategies: the application of a Gradient Reversal Layer (GRL) and a few-shot adaptation technique for a new domain. For the few-shot approach, 5 samples from each activity class were used to fine-tune the pre-trained model for domain adaptation. In the case of the GRL, we modified the DNN architecture to incorporate an additional dense layer following the LSTM output for domain classification, with the GRL applied before this layer to ensure the model prioritizes class-related features and disregards environment-specific features.

\subsection{Results}
\begin{figure}[!h]
    \includegraphics[width=0.47\textwidth]{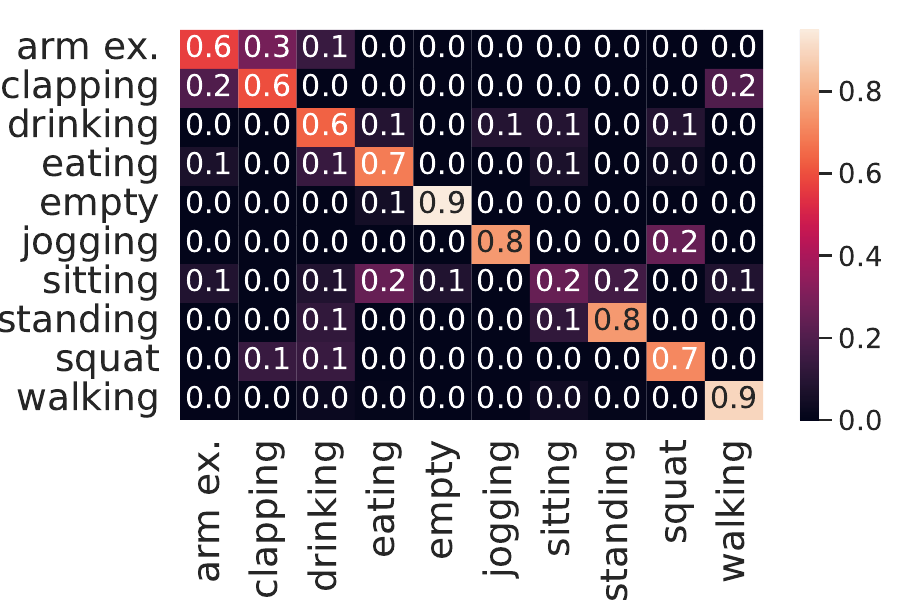}
    \vspace{-10pt}
    \caption{Confusion matrix of activity recognition across all domains.}
    \label{fig:cf}
    \vspace{-5pt}
\end{figure}

In our experimentation, the developed model, tailored for a 10-class classification task, attained a notable overall accuracy of 82\% across various domains, as delineated in the confusion matrix presented in Figure~\ref{fig:cf}, where the rows represent the true class and the columns represent the predicted class. The model demonstrated superior performance in accurately identifying instances of no activity (categorized as 'empty') and the 'walking' activity, which is characterized by significant bodily movements. However, the confusion matrix reveals a specific challenge: approximately 30\% of instances involving 'arm exercise' were misclassified as 'clapping.' This misidentification is attributable to the similarity in hand gestures between these two activities. Furthermore, the model's ability to discern 'sitting down' activities was suboptimal, largely due to the motion resemblance to 'standing up' activities.

\begin{figure}[!h]
    \includegraphics[width=0.47\textwidth]{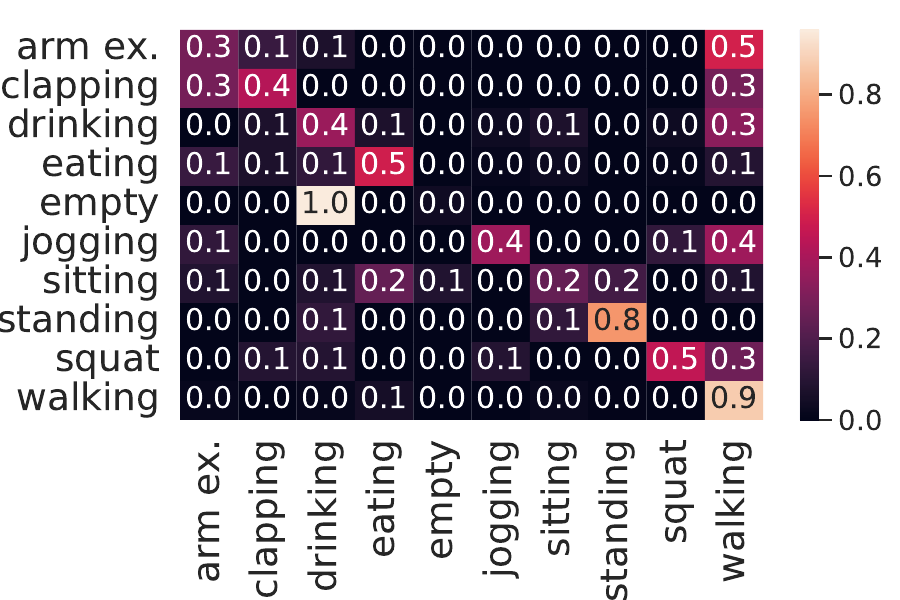}
    \vspace{-10pt}
    \caption{Confusion matrix of activity recognition for unseen domains.}
    \label{fig:cf_e}
    \vspace{-5pt}
\end{figure}

When tested in novel environmental contexts, our model achieves an accuracy of only $42\%$, demonstrating a marked degradation. We show the confusion matrix in Figure~\ref{fig:cf_e}. A key observation from this analysis is the model's inability to accurately classify subtle activities, such as 'drinking,' frequently categorizing them as 'no activity.' Such behavior underscores the model's current limitation in adapting to unfamiliar environments and highlights the necessity for advanced domain adaptation methodologies to enhance its generalization capabilities across diverse environmental conditions.

\begin{figure}[!htb]
    \includegraphics[width=0.47\textwidth]{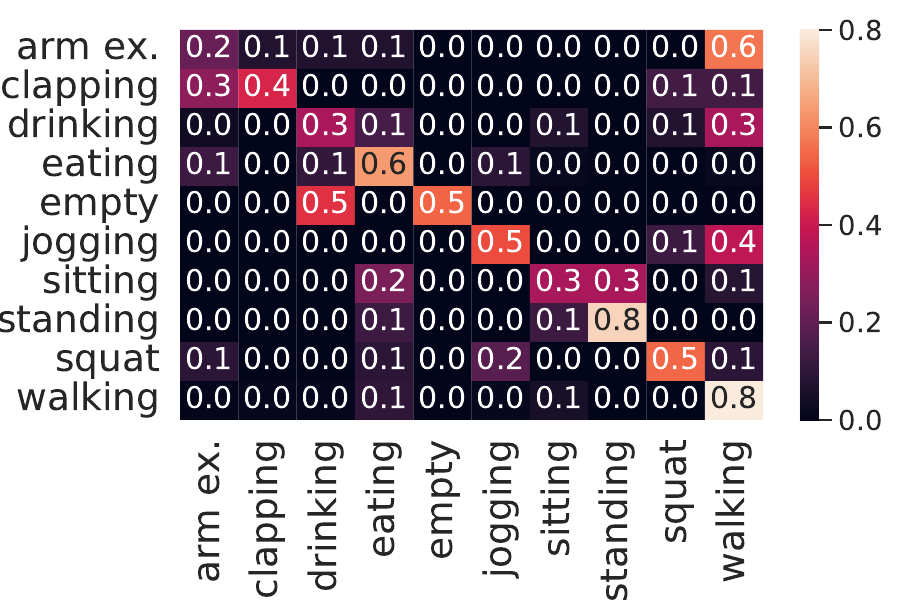}
    \vspace{-10pt}
    \caption{Confusion matrix of activity recognition for unseen domains using domain classification strategy.}
    \label{fig:cf_d}
    \vspace{-5pt}
\end{figure}

In our final set of observations, we noted that the same model, when augmented with an additional domain classification branch preceded by a Gradient Reversal Layer (GRL), attained an improved accuracy of 63\% on datasets derived from the same previously unseen environment. This enhancement is visually represented in the confusion matrix shown in Figure~\ref{fig:cf_d}. A notable shift in the confusion matrix towards a more diagonal pattern was observed, indicating a significant improvement in the model's ability to correctly classify activities, including those involving minute movements. 

\subsection{Effect of Environment}
\begin{figure}[!htb]
    \includegraphics[width=0.47\textwidth]{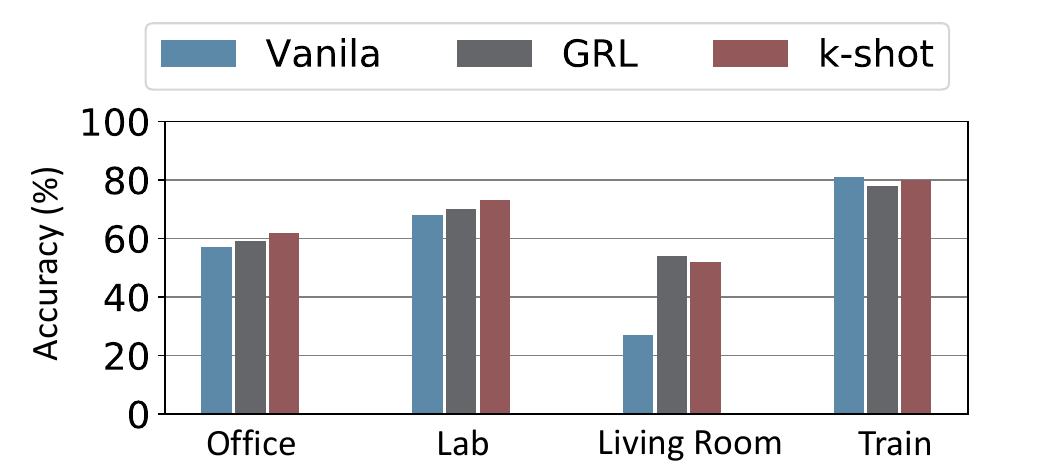}
    \caption{Comparison of different training algorithms acrooss seen and unseen domains.}
    \label{fig:res_env}
\end{figure}

In our research, as detailed in Section~\ref{sec:shift_e}, we evaluated our model's performance across datasets from three unseen environments and compared it with datasets from environments encountered during the training phase. The accuracies were assessed for the base network~(vanila), the enhanced network equipped with an additional domain classification branch and a GRL, and through the implementation of k-shot fine-tuning. Our findings indicate that both domain adaptation strategies yielded improvements in model accuracy for unseen environments, with enhancements ranging between 2\% to 27\%, as shown in Figure~\ref{fig:res_env}. The most significant accuracy increase was observed in the context of an unseen living room environment. Despite these advancements, a notable accuracy gap of 10-30\% persists between the performance in unseen and seen environments. This discrepancy underscores the necessity for more sophisticated approaches to further bridge this divide, suggesting potential avenues for future research in enhancing the model's adaptability and generalization capabilities across varying environmental contexts.

\subsection{Effect of People}
\begin{figure}[!h]
    \includegraphics[width=0.47\textwidth]{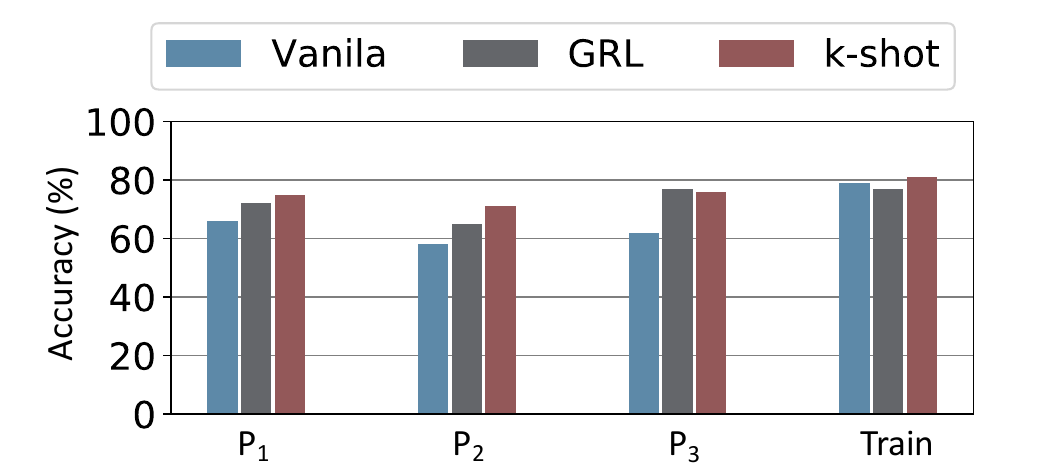}
    \caption{Comparison of different training algorithms acrooss seen and unseen people.}
    \label{fig:res_people}
\end{figure}

Our analysis extends to examining the model's accuracy in recognizing activities performed by different individuals within the same environment, as depicted in Figure~\ref{fig:res_people}. The results indicate that while the model's performance degradation is less pronounced when encountering data from unseen individuals, as compared to entirely new environments, there is still a noticeable accuracy drop of 12-15\% when utilizing the standard (vanilla) training algorithm. However, the implementation of domain adaptation techniques has shown to improve this decline in performance. Specifically, the application of domain adaptation strategies has effectively narrowed the accuracy disparity, reducing it to a range of 2-10\%. 

%% file: tex/06_conclusion.tex
\section{Conclusion and Future Work}

In this paper, we presented the outcomes of our multi-year study, where we deployed WiFi sensing systems in eight distinct environments with 16 participants, after thorough preliminary testing in lab settings. The deployments, ranging from a few hours to two weeks, were meticulously recorded using cameras and wearable devices for accurate ground truth comparison. Our analysis highlighted substantial shifts in data distribution, attributable to variables like individual presence, environmental conditions, and time intervals. We also showed that traditional domain adaptation techniques for handling data shift is not sufficient to fully remove the effect of environment in WiFi sensing. In future, we intend to explore hyper network based approaches to parameterize the environment and model the estimated shift in CSI feature for activity recognition, which can provide a pathway towards developing adaptive WiFi-based sensing systems, particularly in enhancing elderly care.